# Single-shot experimental-numerical twin-image removal in lensless digital holographic microscopy

PIOTR ARCAB*, MIKOŁAJ ROGALSKI, AND MACIEJ TRUSIAK**

*Warsaw University of Technology, Institute of Micromechanics and Photonics, 8 Sw. A. Boboli St., 02-525 Warsaw, Poland*
*\*piotr.arcab.dokt@pw.edu.pl*
*\*\*maciej.trusiak@pw.edu.pl*

**Abstract:** Lensless digital holographic microscopy (LDHM) offers very large field-of-view label-free imaging crucial, e.g., in high-throughput particle tracking and biomedical examination of cells and tissues. Compact layouts promote point-of-case and out-of-laboratory applications. The LDHM, based on the Gabor in-line holographic principle, is inherently spoiled by the twin-image effect, which complicates the quantitative analysis of reconstructed phase and amplitude maps. Popular family of solutions consists of numerical methods, which tend to minimize twin-image upon iterative process based on data redundancy. Additional hologram recordings are needed, and final results heavily depend on the algorithmic parameters, however. In this contribution we present a novel single-shot experimental-numerical twin-image removal technique for LDHM. It leverages two-source off-axis hologram recording deploying simple fiber splitter. Additionally, we introduce a novel phase retrieval numerical algorithm specifically tailored to the acquired holograms, that provides twin-image-free reconstruction without compromising the resolution. We quantitatively and qualitatively verify proposed method employing phase test target and cheek cells biosample. The results demonstrate that the proposed technique enables low-cost, out-of-laboratory LDHM imaging with enhanced precision, achieved through the elimination of twin-image errors. This advancement opens new avenues for more accurate technical and biomedical imaging applications using LDHM, particularly in scenarios where cost-effective and portable imaging solutions are desired.

## 1. Introduction

In the realm of optical microscopy, imaging transparent specimens often poses significant challenges, leading to the rise of quantitative phase imaging technology [1,2]. Especially popular QPI family leverages Digital Holographic Microscopy (DHM) [3–5]. As a well-established technique for quantitative measurements, DHM presents a myriad of solutions by enabling the capture of information pertaining to the complex optical field, which includes both amplitude and phase modulations invoked by the object of study. Holograms within the context of DHM serve to record the intensity of the optical field, which encapsulates two coherent terms: the object wave (+1 term) and the conjugate wave (-1 term); and the incoherent autocorrelation intensity term (0 order). Off-axis DHM systems, based on original Leith and Upatnieks idea [6], are frequently utilized to mitigate the deleterious impact of the conjugate and autocorrelation terms. On-axis methodologies, however, grapple with this issue, commonly referred to as the twin-image effect to be addressed in Gabor holographic schemes [7], e.g., by phase shifting techniques [8,9]. Additional precisely phase-shifted holograms require hardware complications and temporal resolution limitations, however.

In-line Gabor holography is the basis of lensless DHM (LDHM) framework [10–13], which stands out due to its simplicity of hardware, cost-effectiveness, high-throughput and uniquely extensive field of view with very good resolution. The presence of the conjugate image predominantly affects in-line holographic phase measurements [14], therefore numerous experimental and numerical strategies have been formulated to minimize the twin-image to increase the precision of, e.g., quantitative diagnosis based on LDHM bioimaging [15–24].

Let us revisit the numerical origins and minimization schemes of the twin-image effect. The conventional numerical methodology for the reconstruction of the quantitative phase information in LDHM generally requires numerical backpropagation of the intensity distribution of a defocused hologram to the focal plane of the specimen (at appropriate Z distance) [25]. However, this process is impeded by the lack of phase data pertaining to the complex optical field at the camera plane in LDHM. Consequently, when a hologram lacking phase information is backpropagated to the object plane, it results in the superimposition of the specimen's optical field with a digitally generated twin, which is defocused at the inverse Z distance. This interaction leads to the formation of characteristic double-defocused Gabor fringes in the reconstructed in-focus complex optical field - a twin-image phenomenon. This fundamental problem has been the subject of extensive research, resulting in the proposition of several potential solutions. The most prevalent among these is the Gerchberg-Saxton (GS) algorithm [26,27], which necessitates the collection of at least two distinct holograms, acquired with variable wavelengths [28–31] or Z distances [32,33]. Subsequent iterative propagation of the complex-field between the hologram planes enables the retrieval of the phase factor of the complex field in the hologram plane, with appropriate recorded intensity constraints. However, this approach has a significant drawback in that it requires the efficient collection of several distinct holograms, thereby adding to the complexity of both hardware and software systems. From an economic perspective, this process also extends the measurement duration, effectively negating one of DIHM's primary advantages - its simplicity and robustness. Similar problems are significantly increased when entire training set needs to be collected and labelled for neural network solutions [34]. When considering the removal of the twin-image effect using a single hologram, the GS algorithm may again be employed. In this instance, the hologram is propagated between the object and hologram planes, with the implementation of appropriate constraints in the object plane [35,36]. Nevertheless, this solution demands extensive a priori knowledge about the sample, often necessitating the masking of object regions, and typically lacks the effectiveness of the multi-hologram GS method as it operates without data-multiplexing. An alternative group of proposed solutions includes the regularization approaches [37,38]. Here, the twin-image at the object plane is iteratively diffused using a range of norms (e.g., TV), while the sharp features of the object remain unaltered. However, this approach also necessitates some degree of a priori information regarding the measured object and typically requires a substantial number of iterations, making it a time-consuming solution.

On the somewhat opposite side of the court, we have the experimental techniques to remove or minimize the twin-image effect. A capable family of setup solutions is composed of so-called lensless off-axis schemes, generating a carrier fringe pattern to be analyzed upon Fourier transform with deterministic phase demodulation, hence limited or even removed twin-image (which originates, as mention in previous paragraph, from the lack of phase values within the hologram plane). Rostykus and Moser proposed very clever lensless off-axis microscopy [39], deploying VCSEL laser source and especially tailored beam dividing elements. Although very elegant, straightforward and efficient, those solution required manufacturing of a very specific beamsplitter element, based on gratings with fixed period, thus working for a given setup and sample geometry and effective hologram carrier frequency within a limited range of adjustments. Serabyn et al. described a double-pinhol setup for lensless off-axis holographic microscopy in total-shear regime [40]. Double-pinhol setup was recently improved via additional of GRIN lens [41]. Due to limited geometry settings, the lensless imaging parameters, e.g., resolution [40,41] and signal to noise ratio of numerically refocused phase map, can be easily affected [39–41].

In this research paper, we propose an easy-to-implement off-axis transmission LDHM method that employs the off-the-shelf fiber-coupler. Our proposed system amalgamates the advantages of LDHM in terms of simplicity with the benefits of off-axis DHM, predominantly the elimination of the twin-image effect during the reconstruction process upon novel tailored numerical scheme involving off-axis phase demodulation and in-line complex field backpropagation. Novel experimental setup and numerical reconstruction framework are described and validated in next four chapters. Discussion and conclusion sections finalize the paper.

## 2. Single-shot TIR-LDHM experimental setup description

Figure 1 depicts a novel Twin-Image Removed Lensless Digital Holographic Microscopy (TIR-LDHM) system incorporating an off-axis reference beam on top of classical in-line Gabor configuration with common-path scattered and unscattered beams. To ensure simplicity in the experimental setup, a laser (CNI Lasers MGL-FN-561-20mW, $\lambda = 561$ nm, FWHM = 47 pm) was utilized as the light source and coupled to a single mode narrowband fiber coupler (Thorlabs TN560R5F1). The fiber coupler's alignment involved positioning one of its connectors in line with the camera (ALVIUM Camera 1800 U-2050 m mono Bareboard, pixel size 2.4 × 2.4 μm, 5496 × 3672 pixels, FOV = 116 mm²) to generate classical Gabor in-line lensless hologram. The second connector, placed in a reference off-axis fashion (see Fig. 1), served two purposes: (1) avoiding sample illumination, which could introduce additional errors, and (2) achieving the appropriate carrier frequency in accordance with Nyquist criterion. The separation between two point sources $x$ can be calculated basing on Eq (1):

$$x = \tan\left[\sin^{-1}\left(\frac{\lambda}{d_f}\right)\right] * D, \qquad (1)$$

where $\lambda$ denotes the light wavelength, $d_f$ stands for fringes period and $D$ is the distance between detector and the connector placed on-axis. To obtain the optimal frequency separation of the background and carrier fringes information, the $d_f$ should be equal $2 \cdot d_x \cdot \sqrt{2}$ (for 45 deg fringes orientation), where $d_x$ is the camera pixel size. In our proof-of-concept configuration, the sample was positioned at the center of the illuminating beam, and the magnification of the setup was approximately M=2. The entire setup had a length of $D$=100 mm, allowing for easy control of the carrier frequency, reconfigurable with respect to the pixel size (easily implementable with various cameras), by adjusting the second connector. Adjustment flexibility induced larger dimensions of the setup, especially in comparison to very compact devices [39], however the miniaturization of the experimental architecture is not our goal in this proof-of-concept contribution reporting novel lensless holographic technique called TIR-LDHM. For our system the transverse length between fiber connectors $x$ is approximately equal to 8,3 mm.

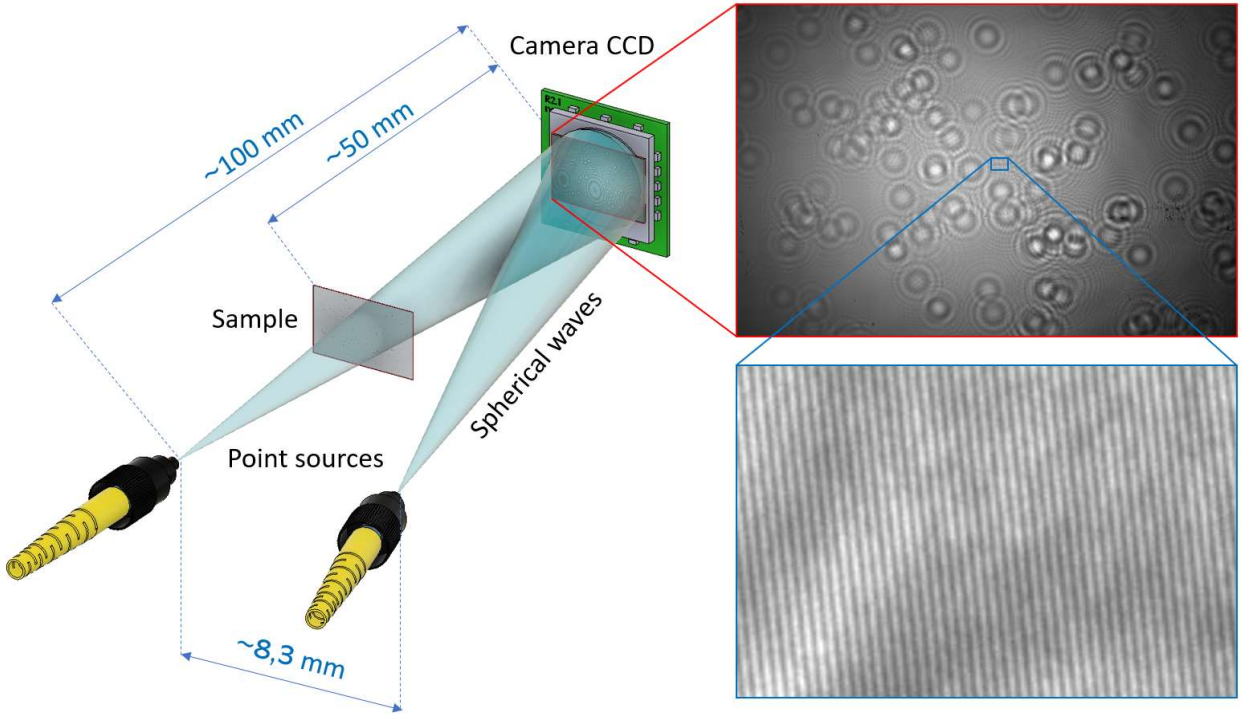

Fig. 1. Scheme of proof-of-concept lensless digital holographic microscope with two point sources illumination, along with exemplary recorded hologram.

### 3. Tailored numerical algorithm for single-shot TIR-LDHM

Figure 2 illustrates a novel single-shot numerical reconstruction method supporting TIR-LDHM twin-image elimination. It deploys multiplexed in-line and off-axis holograms recorded in the presented system of lensless holographic microscopy. To separate the object and conjugate information in the frequency spectrum, dense carrier fringes were introduced into Gabor's hologram. This technique is based on the principles of Fourier transform [42]. Reconstruction process commences with the initial step of performing the Fourier transform of the input hologram. Subsequently, a mask is generated to isolate one of the cross-correlation peaks in the Fourier spectrum, which contains crucial coherent information about the complex optical field stored in the carrier fringes. To create this mask, we employ the OTSU method to automatically binarize the logarithm of the Fourier spectrum using a determined threshold. This binarized image then undergoes a morphological erosion operation repeatedly until it contains less than 5% of "1" values, to eliminate the small binarized groups, ensuring that the binarized side peaks are disconnected from the central peak. After that, morphological dilation operation is applied to restore the original mask groups dimensions. The largest binarized area outside the spectrum center is selected as the final binary mask ("$M_p$" in Fig. 2), which is further smoothed using Gaussian filtering with a standard deviation of 10. Based on "$M_p$", a second mask ("$M_a$" in Fig. 2) is created by inverting the sum of "$M_p$" with a copy of itself, symmetrically reflected relative to the image center (i.e., rotated by 180 degrees).

Next, the hologram Fourier spectrum is filtered using "$M_p$" mask. The filtered portion of the spectrum is then centered, and the inverse Fourier transform is applied to recover the complex optical field at the camera plane ("$C_{cam}$" in Fig. 2). Simultaneously, the hologram Fourier spectrum is also filtered using "$M_a$" mask, and the inverse Fourier transform is performed to retrieve the amplitude part of the optical field in the camera plane ("$A_{cam}$" in Fig. 2) without the carrier fringes factor. The phase component of "$C_{cam}$" is subsequently unwrapped and fitted to a plane ("$P_{cam}$" in Fig. 2) to eliminate the spherical phase factors originating from both illumination sources. The fitting to the plane is achieved by subtracting the unwrapped phase from its Gaussian-blurred version (with a standard deviation of 60).

The final optical field at the camera plane is constructed as $OF_{cam} = A_{cam} \cdot e^{iP_{cam}}$. It is worth noting that instead of employing additional operations to retrieve "$A_{cam}$", one can simply utilize the amplitude component of "$C_{cam}$". However, since "$A_{cam}$" contains a significantly higher number of spatial frequencies compared to "$C_{cam}$" (as evident from the masks "$M_a$" and "$M_p$" used to derive each of them), employing "$A_{cam}$" allows for a higher resolution reconstruction than with a pure "$C_{cam}$" reconstruction – we show this in Fig. 3 comparing both methods' results presented in Fig. 3(a) and Fig. 3(b).

Moving forward, "$OF_{cam}$" is backpropagated to the object plane using the angular spectrum (AS) method [43–46], resulting in twin-image-free complex optical field reconstruction ("$OF_{obj}^{(1)}$" in Fig. 2). Nevertheless, due to the limited number of spatial frequencies present in the phase component of "$OF_{cam}$" compared to "$A_{cam}$", the achieved resolution may still be inferior to that obtained by directly backpropagating "$A_{cam}$" to the object plane ("$OF_{obj}^{(2)}$" in Fig. 2). To address this compromise, we introduce an additional mask ("$M_{obj}$" in Fig. 2) by shifting the "$M_p$" to the image center, and combine the low frequencies of "$OF_{obj}^{(1)}$" with the high frequencies of "$OF_{obj}^{(2)}$" to produce the final reconstruction result ("$OF_{obj}$" in Fig. 2), which ensures a twin-image-free reconstruction without resolution loss (compare results presented in Fig. 3(b) and Fig. 3(c)):

$$OF_{obj} = F^{-1}\left(F\left(OF_{obj}^{(1)}\right) \cdot M_{obj} + F\left(OF_{obj}^{(2)}\right) \cdot \left(1 - M_{obj}\right)\right), \qquad (2)$$

where F and F-1 stands for Fourier transform and inverse Fourier transform, respectively.

It is important to emphasize that, for the numerical propagation between the camera and sample plane, we have employed the AS method (as mentioned before), a well-established technique widely used for optical field propagation under the paraxial approximation, which is applicable to our specific case. To achieve successful reconstruction of in-focus information using the AS method, it is essential to accurately determine the distance between the sample and the camera. For this purpose, we have utilized the DarkFocus [47] metric, a representative of autofocusing algorithms [48–54], which are of paramount significance in numerical focusing and play a crucial role in LDHM. However, it is to be noted that, in situations involving larger magnifications and spherical wavefronts, appropriate propagation routines should be implemented, as outlined in prior works [25,55,56].

Another important thing to underline is that the proposed algorithm performance strongly depends on the generation of Fourier spectrum masks and associated with this process morphological operations. Theoretically, in the case of generating too narrow mask, some frequency information will be cut-off, resulting in the resolution loss. On the other hand, mask with larger dimensions should ensure that all frequency components are employed in the reconstruction, however, it may result in higher noise in the reconstruction. To investigate the effect of different mask dimensions on the signal-to-noise ratio, an additional parameter "d" was introduced. This parameter represents the number of extra pixels added to each edge of the mask (in other words, the dilation operation with disk kernel of "d" radius was performed on binary mask "$M_p$" before Gaussian smoothing). By varying the "d" parameter, the goal is to identify the mask "$M_p$" (and so the "$M_a$" and "$M_{obj}$" which shape is defined by "$M_p$") dimensions that result in the highest signal-to-noise ratio, optimizing the reconstruction process. Figure 4(left) shows the mean of the spectrum depicted above the chart in the vertical direction, and cross sections of masks with different "d" parameter. The standard deviation (STD) from the reconstructed object-free are was used to quantify noise level [57,58], and the results are presented in Figure 4 on the righthand side. As the "d" parameter is increased, it leads to a corresponding increase in noise. Despite increasing the noise there is no visible resolution increasement as shown in inserts presenting biological sample (cheek cells) with detailed features.

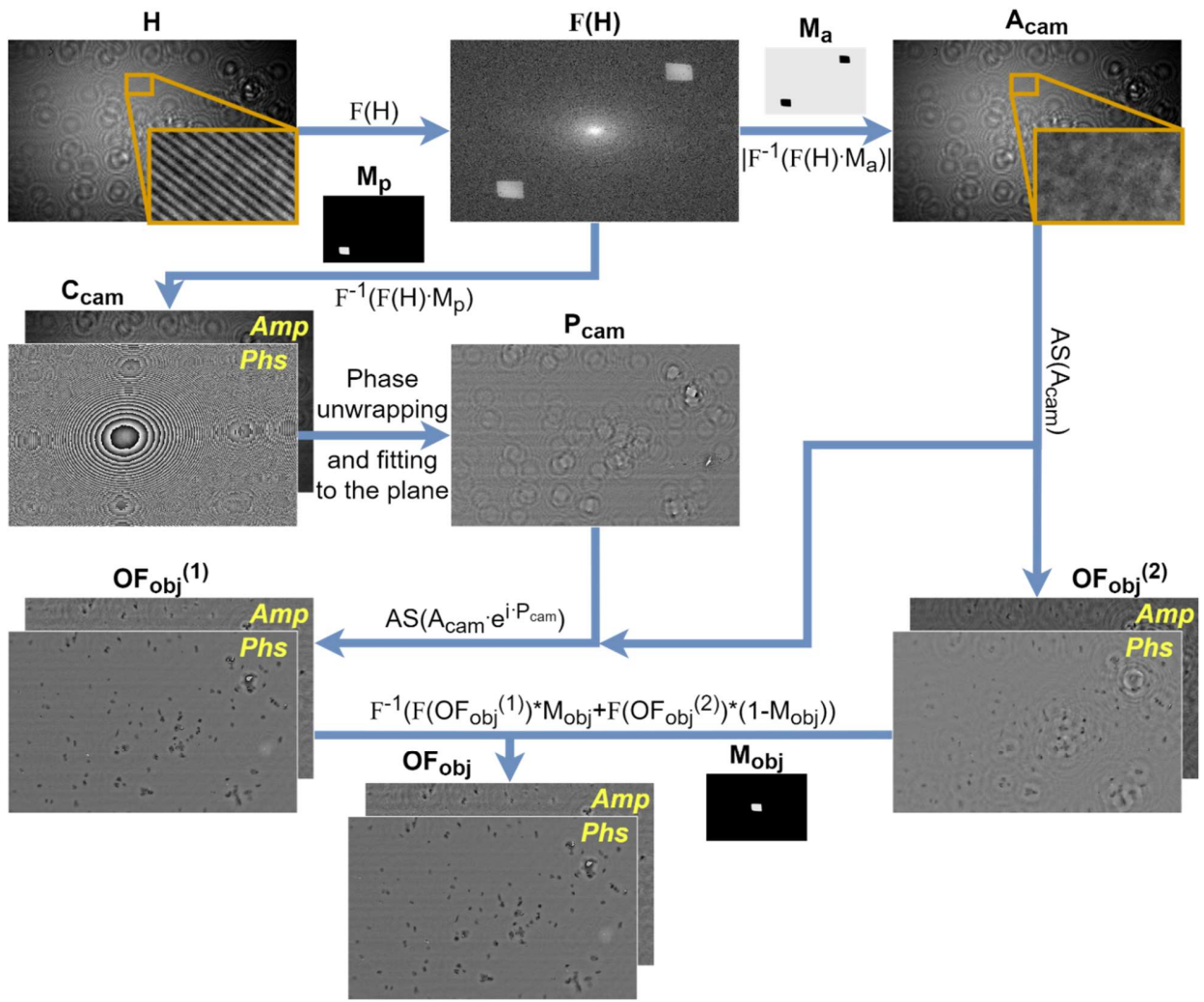

Fig. 2. Diagram of hologram (H) processing registered in setup for twin-image elimination. F – Fourier transform. F$^{-1}$ – inverse Fourier transform. $M_p$ and $M_a$ – masks applied on frequency spectrum to retrieve phase ($P_{cam}$) and amplitude without carrier frequency ($A_{cam}$) from registered hologram, respectively. $OF_{obj}$ – reconstructed optical fields at the object plane. AS – angular spectrum propagation method.

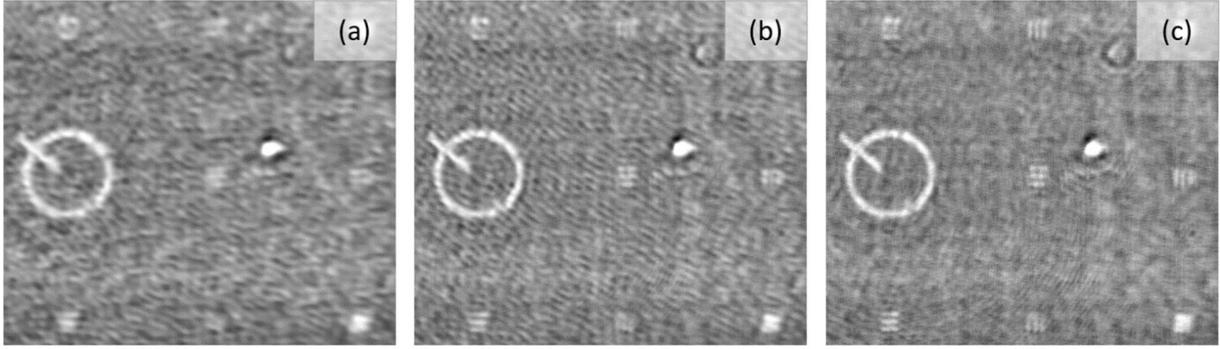

Fig. 3. Comparison of resolution in reconstructions (a) – $|C_{cam}| \cdot e^{iP_{cam}}$ backpropagated to the object plane, (b) $OF_{obj}^{(1)}$ and (c) $OF_{obj}$.

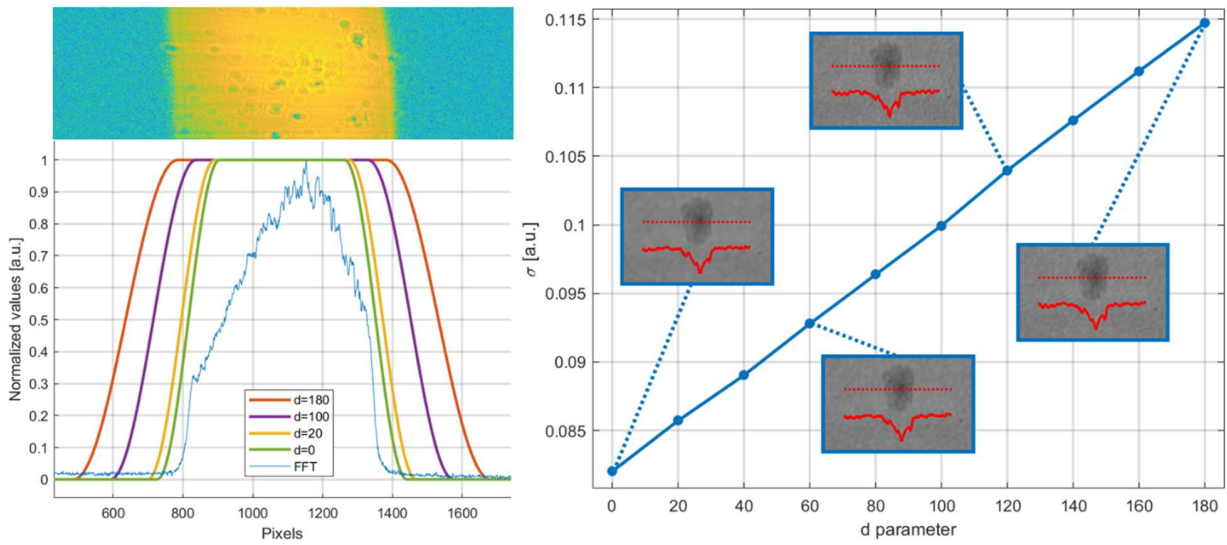

Fig. 4. The image on the lefthand side displays a comparison of mask cross-sections with varying values of the parameter "d" and the mean in vertical direction of spectrum +1 term, shown at the top, named FFT. The righthand side image depicts the noise, calculated using the STD from an area without the object, corresponding to different values of the parameter "d." In some instances of the parameter "d," a biological sample was illustrated to demonstrate that the resolution remains constant despite the changes in the "d" parameter and increasing noise.

## 4. Experimental verification: phase test target examination

To evaluate the effectiveness of the proposed method in terms of the spatial resolution and reduction of twin images, we conducted comparative experiments using a standard lensless holographic microscopy system. A custom-made phase test target (Lyncée Tec, Boroflat 33 glass, 125±5 nm height) was used for these experiments. In the experiment, we first recorded a hologram using the classical single-path LDHM system as the baseline. Additionally, we captured a second hologram with a slight displacement of the camera to enable the application of the well-known and commonly used Gerchberg-Saxton (GS) algorithm in a multi-height fashion [32,33], which aims to minimize the effects of twin- image. Figure 5 illustrates the reconstructions obtained from the different methods: (a) hologram recorded and reconstructed in standard LDHM, (c) four-hologram GS algorithm in standard LDHM, and (b) hologram recorded and demodulated via novel TIR-LDHM method (deploying a fiber coupler like discussed in Fig. 1). Enlarged areas with the finest visible features are shown in blue and orange boxes, respectively, allowing for a clear comparison of the reconstructions. In Fig. 5(a), twin-image fringes are significantly pronounced since it represents a straightforward reconstruction without any additional processing. In Fig. 5(c), twin-image fringes are considerably reduced; however, low frequency twin-image artefacts are still visible as they are the most cumbersome to eliminate with GS (low frequency twin-image components are similar in all recorded holograms). Figure 5(b) exhibits the most notable distinctions compared to Fig. 5(a). Specifically, the vertical and horizontal lines resulting from twin-image effect have been nearly completely eliminated, leaving only some speckle noise as the remaining artifact (speckles are present in all reconstructions, twin-image removal highlights them, however). It is important to note that the resolution remains the same in all three methods, in both the vertical and horizontal directions.

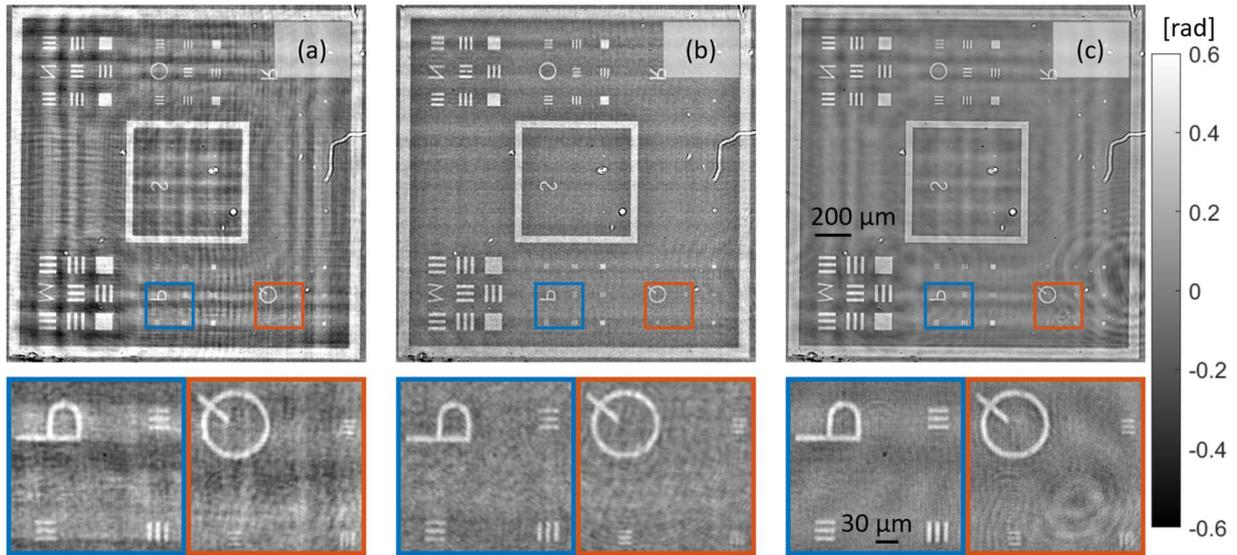

Fig. 5. Comparison of reconstruction of standard LDHM hologram (a), proposed TIR-LDHM method (b) and GS 4-hologram iterative approach (c). Blue and orange boxes mark the enlarged areas of the finest elements.

## 5. Experimental corroboration: live cheek cells examination

The proposed method has been validated using unfixed human cheek cells as a challenging biosample. In Fig. 6(a), the reconstruction of the hologram recorded in LDHM is presented, while in Fig. 6(b), the tailored reconstruction of the novel TIR-LDHM hologram is shown. Figure 6 includes enlarged areas with a cross-sections that contain fine bio-features, specifically the nucleus of the cheek cell. When observing the hologram reconstructions within a large field of view (FOV), it becomes evident that there is a significant improvement in reducing twin-image fringes. The impact of twin-image reduction is clearly visible, indicating the effectiveness of the proposed method in enhancing the quality of the reconstructed holographic images across a wider FOV. Furthermore, the calculated STD value from area marked by yellow box representing noise in Fig. 6(a) amounts to 0.0952 rad, whereas in Fig. 6(b), it is reduced to 0.0817 rad.

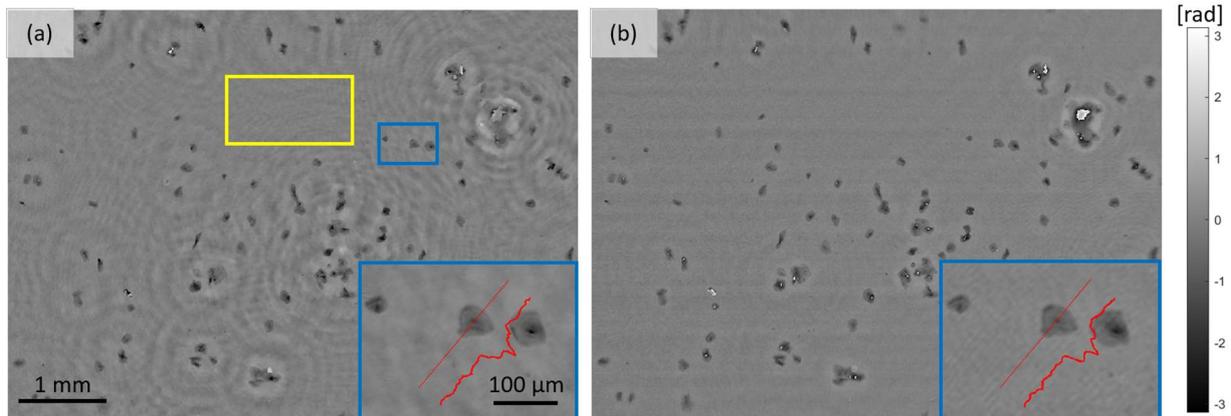

Fig. 6. Evaluation of proposed method with additional transmission beam for biological samples: human cheek cells. (a) Reconstruction of hologram recorded in standard LDHM, (b) Reconstruction of the hologram recorded in LDHM system with additional transmission beam. Yellow box in (a) marks the area which was used to calculate the noise.

## 6. Discussion

Proposed TIR-LDHM addresses the serious challenges posed by the twin-image phenomenon, which arises due to the lack of phase data in the hologram plane, hindering the reconstruction of complex optical fields. Our system overcomes this issue and offers significant advantages over existing techniques in terms of simplicity, hardware requirements and speed (number of holograms required). The key experimental element of our proposed method is the incorporation of an off-axis reference beam alongside the classical in-line Gabor configuration in a LDHM setup. The pivotal numerical reconstruction algorithm supporting TIR-LDHM innovatively involves the processing of multiplexed in-line/off-axis hologram to separate the object and conjugate information in the frequency spectrum. This is based on automatically generating Fourier spectrum masks and employing morphological operations to ensure the extraction of essential coherent information while eliminating unwanted artifacts. By carefully designing and combining these masks, we achieve a twin-image-free reconstruction without sacrificing the resolution, which we corroborate experimentally.

Comparative experiments using a standard LDHM system demonstrate the effectiveness of our proposed TIR-LDHM method. Reconstructions obtained through our novel approach exhibit a significant reduction in twin-image fringes, leaving only minor speckle noise as the remaining artifact, over-highlighted by the lack of twin-image components. In the Gerchberg-Saxton numerical iterative method, speckle component is averaged over multiple frames, thus is of more gentle influence than in single-shot TIR-LDHM. Nonetheless, speckle noise in TIR-LDHM is to be possibly minimized in the future works, e.g., via physical (coherence) or numerical (post-processing) alterations.

Importantly, the spatial resolution remains the same as in standard LDHM, ensuring the preservation of fine details in the reconstructed images. We have also validated our method using unfixed human cheek cells as a challenging biosample. The reconstructed images showcase a notable improvement in reducing twin-image fringes, showcasing the potential of our proposed TIR-LDHM approach for enhancing the quality of holographic bio-imaging across a larger field of view in point-of-care conditions (once converted from lab system into a lensless prototype). The proposed TIR-LDHM system offers several advantages over existing techniques for twin-image removal. By utilizing off-the-shelf fiber-couplers and incorporating off-axis reference beams, the experimental setup becomes simple, cost-effective, and suitable for high-throughput imaging. Moreover, the elimination of the twin-image effect is achieved using a single-shot hologram, reducing the complexity of both hardware and software systems, and avoiding the need for extensive data-multiplexing or a priori knowledge about the sample, and maximizing the temporal resolution of the imaging.

Future work could involve further optimization of the system and algorithm parameters to maximize the signal-to-noise ratio while exploring its potential applications in real-time imaging and diagnostics. We further discuss some possible ways for advancements. (1) Our setup requires the separation of object and reference beams - the sample thus must be far from the camera, which can lead to lower overall resolution compared to classical LDHM with the sample close to the camera, due to optical magnification around 2x in TIR-LDHM and 1x in classical LDHM. Care must be taken when assessing the resolution, however, as it is limited by a number of factors, e.g., coherence, sensor size, wavelength etc., but mainly by the pixel size in classical LDHM. Pixel size is effectively decreased using 2x magnification in TIR-LDHM, enhancing the hologram spatial sampling rate. One possible future solution to the mentioned beam separation problem could be placing the sample close to the camera and working in the regime of shearing interferometry with appropriate numerical solutions. (2) It is to be showcased that two spherical beams are superimposed in our TIR-LDHM method, thus those two spherical phase components need to be removed during the reconstruction. It should be noted that we handled this well within the proposed novel algorithm, however. (3) Higher spatio-temporal coherence of the light source is generally required in TIR-LDHM than in the classical LDHM because not only should the axial "Gabor" beams be mutually coherent, but also the off-axis reference one should interfere with the axial beams.

## 7. Conclusions

In this study, we have presented a novel technique for mitigating the twin-image effect in lensless digital holographic microscopy (LDHM). By leveraging a two-source off-axis hologram recording setup and a tailored numerical algorithm, we were able to calculate the on-axis phase and amplitude distribution in the camera plane and perform twin-image-free complex field propagation using the angular spectrum method. Our experimental and numerical verification using a phase test target and cheek cells biosample demonstrated the effectiveness of the proposed technique. The removal of twin-image artifacts in LDHM is of paramount importance for achieving accurate and reliable quantitative analysis of reconstructed phase and amplitude maps. Compared to existing methods, our single-shot approach eliminates the need for additional hologram recordings and reduces the dependence on algorithmic parameters. This not only simplifies the imaging process but also increases precision and makes LDHM more accessible for low-cost out-of-laboratory applications. The reported technique opens up new possibilities for high-throughput easy-to-implement quantitative phase imaging for particle tracking, biomedical examination of cells and tissues, and other label-free imaging applications in various research fields, including biology, medicine, and materials science. With the twin-image effect effectively addressed, researchers and practitioners can benefit from the larger field-of-view and improved imaging quality provided by LDHM. Future work could involve further optimization of the system and algorithm parameters to maximize the resolution and signal-to-noise ratio while exploring its potential applications in real-time imaging and diagnostics.

**Funding.** This work has been funded by the National Science Center Poland (SONATA 2020/39/D/ST7/03236). M.R. is supported by the Foundation for Polish Science (FNP start program).

**Disclosures.** The authors declare no conflicts of interest.

**Acknowledgements.** The authors would like to thank Emilia Wdowiak for fruitful discussions and support.

**Data availability.** Data underlying the results presented in this paper are not publicly available at this time but may be obtained from the authors upon reasonable request.

## References

1. Y. K. Park, C. Depeursinge, and G. Popescu, "Quantitative phase imaging in biomedicine," Nat. Photonics **12**, 578–589 (2018).
2. G. Popescu, *Quantitative Phase Imaging of Cells and Tissues*, 1st Edition (McGraw-Hill Education, 2011).

3. B. Kemper and G. Von Bally, "Digital holographic microscopy for live cell applications and technical inspection," Appl. Opt. **47**, A52-A61 (2008).
4. Y. Cotte, F. Toy, P. Jourdain, N. Pavillon, D. Boss, P. Magistretti, P. Marquet, and C. Depeursinge, "Marker-free phase nanoscopy," Nat. Photonics **7**, 113–117 (2013).
5. U. Schnars and W. Jueptner, *Digital Holography* (Springer-Verlag, 2005).
6. E. N. Leith and J. Upatnieks, "Microscopy by Wavefront Reconstruction," J. Opt. Soc. Am. **55**, 569_1-570 (1965).
7. D. Gabor, "A new microscopic principle," Nature **161**, 777–778 (1948).
8. I. Yamaguchi and T. Zhang, "Phase-shifting digital holography," Opt. Lett. **22**, 1268-1270 (1997).
9. V. Micó, J. García, Z. Zalevsky, and B. Javidi, "Phase-shifting Gabor holography," Opt. Lett. **34**, 1492-1494 (2009).
10. Y. Zhang, Y. Shin, K. Sung, S. Yang, H. Chen, H. Wang, D. Teng, Y. Rivenson, R. P. Kulkarni, and A. Ozcan, "3D imaging of optically cleared tissue using a simplified CLARITY method and on-chip microscopy," Sci. Adv. **3**, e1700553 (2017).
11. A. Greenbaum, W. Luo, T. W. Su, Z. Göröcs, L. Xue, S. O. Isikman, A. F. Coskun, O. Mudanyali, and A. Ozcan, "Imaging without lenses: Achievements and remaining challenges of wide-field on-chip microscopy," Nat. Methods **9**, 889-895 (2012).
12. A. Ozcan and E. McLeod, "Lensless Imaging and Sensing," Annu. Rev. Biomed. Eng. **18**, 77–102 (2016).
13. E. McLeod and A. Ozcan, "Unconventional methods of imaging: Computational microscopy and compact implementations," Reports Prog. Phys. **79**, 076001 (2016).
14. J. Garcia-Sucerquia, W. Xu, S. K. Jericho, P. Klages, M. H. Jericho, and H. J. Kreuzer, "Digital in-line holographic microscopy," Appl. Opt. **45**, 836-850 (2006).
15. W. Xu, M. H. Jericho, I. A. Meinertzhagen, and H. J. Kreuzer, "Digital in-line holography for biological applications," Proc. Natl. Acad. Sci. U. S. A. **98**, 11301-11305 (2001).
16. S. K. Jericho, P. Klages, J. Nadeau, E. M. Dumas, M. H. Jericho, and H. J. Kreuzer, "In-line digital holographic microscopy for terrestrial and exobiological research," Planet. Space Sci. **58**, 701-705 (2010).
17. T.-W. Su, L. Xue, and A. Ozcan, "High-throughput lensfree 3D tracking of human sperms reveals rare statistics of helical trajectories," Proc. Natl. Acad. Sci. **109**, 16018–16022 (2012).
18. J. P. Ryle, S. McDonnell, B. Glennon, and J. T. Sheridan, "Calibration of a digital in-line holographic microscopy system: Depth of focus and bioprocess analysis," Appl. Opt. **52**, C78-87 (2013).
19. R. Corman, W. Boutu, A. Campalans, P. Radicella, J. Duarte, M. Kholodtsova, L. Bally-Cuif, N. Dray, F. Harms, G. Dovillaire, S. Bucourt, and H. Merdji, "Lensless microscopy platform for single cell and tissue visualization," Biomed. Opt. Express **11**, 2806-2817 (2020).
20. I. Pushkarsky, Y. Lyb, W. Weaver, T. W. Su, O. Mudanyali, A. Ozcan, and D. Di Carlo, "Automated single-cell motility analysis on a chip using lensfree microscopy," Sci. Rep. **4**, 4717 (2014).
21. S. Amann, M. von Witzleben, and S. Breuer, "3D-printable portable open-source platform for low-cost lens-less holographic cellular imaging," Sci. Rep. **9**, 11260 (2019).
22. Y. Wu and A. Ozcan, "Lensless digital holographic microscopy and its applications in biomedicine and environmental monitoring," Methods **136**, 4–16 (2018).
23. H. Zhu, S. O. Isikman, O. Mudanyali, A. Greenbaum, and A. Ozcan, "Optical imaging techniques for point-of-care diagnostics," Lab Chip **13**, 51-67 (2013).
24. V. Boominathan, J. T. Robinson, L. Waller, and A. Veeraraghavan, "Recent advances in lensless imaging," Optica **9**, 1-16 (2022).
25. T. Latychevskaia and H.-W. Fink, "Practical algorithms for simulation and reconstruction of digital in-line holograms," Appl. Opt. **54**, 2424-2434 (2015).
26. R. W. Gerchberg, "A practical algorithm for the determination of phase from image and diffraction plane pictures," Optik (Stuttg). **35**, 237–246 (1972).
27. J. R. Fienup, "Phase retrieval algorithms: a comparison," Appl. Opt. **21**, 2758-2769 (1982).

28. C. Zuo, J. Sun, J. Zhang, Y. Hu, and Q. Chen, "Lensless phase microscopy and diffraction tomography with multi-angle and multi-wavelength illuminations using a LED matrix," Opt. Express **23**, 14314-14328 (2015).
29. M. Sanz, J. Á. Picazo-Bueno, L. Granero, J. García, and V. Micó, "Four channels multi-illumination single-holographic-exposure lensless Fresnel (MISHELF) microscopy," Opt. Lasers Eng. **110**, 341-347 (2018).
30. L. Herve, O. Cioni, P. Blandin, F. Navarro, M. Menneteau, T. Bordy, S. Morales, and C. Allier, "Multispectral total-variation reconstruction applied to lens-free microscopy," Biomed. Opt. Express **9**, 5828-5836 (2018).
31. V. Micó, M. Rogalski, J. Á. Picazo-Bueno, and M. Trusiak, "Single-shot wavelength-multiplexed phase microscopy under Gabor regime in a regular microscope embodiment," Sci. Rep. **13**, 4257 (2023).
32. A. Greenbaum, U. Sikora, and A. Ozcan, "Field-portable wide-field microscopy of dense samples using multi-height pixel super-resolution based lensfree imaging," Lab Chip **12**, 1242-1245 (2012).
33. Y. Rivenson, Y. Wu, H. Wang, Y. Zhang, A. Feizi, and A. Ozcan, "Sparsity-based multi-height phase recovery in holographic microscopy," Sci. Rep. **6**, 37862 (2016).
34. Y. Rivenson, Y. Wu, and A. Ozcan, "Deep learning in holography and coherent imaging," Light Sci. Appl. **8**, 85 (2019).
35. L. Denis, C. Fournier, T. Fournel, and C. Ducottet, "Numerical suppression of the twin image in in-line holography of a volume of micro-objects," Meas. Sci. Technol. **19**, 074004 (2008).
36. S. M. F. Raupach, "Cascaded adaptive-mask algorithm for twin-image removal and its application to digital holograms of ice crystals," Appl. Opt. **48**, 287-301 (2009).
37. W. Zhang, L. Cao, D. J. Brady, H. Zhang, J. Cang, H. Zhang, and G. Jin, "Twin-Image-Free Holography: A Compressive Sensing Approach," Phys. Rev. Lett. **121**, 093902 (2018).
38. A. S. Galande, H. P. R. Gurram, A. P. Kamireddy, V. S. Venkatapuram, Q. Hasan, and R. John, "Quantitative phase imaging of biological cells using lensless inline holographic microscopy through sparsity-assisted iterative phase retrieval algorithm," J. Appl. Phys. **132**, 243102 (2022).
39. M. Rostykus and C. Moser, "Compact lensless off-axis transmission digital holographic microscope," Opt. Express **25**, 16652-16659 (2017).
40. E. Serabyn, K. Liewer, and J. K. Wallace, "Resolution optimization of an off-axis lensless digital holographic microscope," Appl. Opt. **57**, A172-A180 (2018).
41. A. A. Khorshad and N. Devaney, "GRIN-lens-based in-line digital holographic microscopy," Appl. Opt. **62**, D131-D137 (2023).
42. M. Takeda, H. Ina, and S. Kobayashi, "Fourier-transform method of fringe-pattern analysis for computer-based topography and interferometry," J. Opt. Soc. Am. **72**, 156-160 (1982).
43. F. Shen and A. Wang, "Fast-Fourier-transform based numerical integration method for the Rayleigh-Sommerfeld diffraction formula," Appl. Opt. **45**, 1102-1110 (2006).
44. K. Matsushima and T. Shimobaba, "Band-limited angular spectrum method for numerical simulation of free-space propagation in far and near fields," Opt. Express **17**, 19662-19673 (2009).
45. T. Kozacki, K. Falaggis, and M. Kujawinska, "Computation of diffracted fields for the case of high numerical aperture using the angular spectrum method," Appl. Opt. **51**, 7080-7088 (2012).
46. T. Kozacki and K. Falaggis, "Angular spectrum-based wave-propagation method with compact space bandwidth for large propagation distances," Opt. Lett. **40**, 3420-3423 (2015).
47. M. Trusiak, J. A. Picazo-Bueno, P. Zdankowski, and V. Micó, "DarkFocus: numerical autofocusing in digital in-line holographic microscopy using variance of computational dark-field gradient," Opt. Lasers Eng. **134**, 106195 (2020).
48. P. Langehanenberg, G. von Bally, and B. Kemper, "Autofocusing in digital holographic microscopy," 3D Res. **2**, 4 (2011).
49. F. Dubois, A. El Mallahi, J. Dohet-Eraly, and C. Yourassowsky, "Refocus criterion for both phase and amplitude objects in digital holographic microscopy," Opt. Lett. **39**, 4286-4289 (2014).

50. P. Memmolo, M. Paturzo, B. Javidi, P. A. Netti, and P. Ferraro, "Refocusing criterion via sparsity measurements in digital holography," Opt. Lett. **39**, 4719-4722 (2014).
51. C. A. Trujillo and J. Garcia-Sucerquia, "Automatic method for focusing biological specimens in digital lensless holographic microscopy," Opt. Lett. **39**, 2569-2572 (2014).
52. Y. Zhang, H. Wang, Y. Wu, M. Tamamitsu, and A. Ozcan, "Edge sparsity criterion for robust holographic autofocusing," Opt. Lett. **42**, 3824-3827 (2017).
53. Y. Wu, Y. Rivenson, Y. Zhang, Z. Wei, H. Günaydin, X. Lin, and A. Ozcan, "Extended depth-of-field in holographic imaging using deep-learning-based autofocusing and phase recovery," Optica **5**, 704-710 (2018).
54. H. Wang, M. Lyu, and G. Situ, "eHoloNet: a learning-based end-to-end approach for in-line digital holographic reconstruction," Opt. Express **26**, 22603-22614 (2018).
55. K. M. Molony, B. M. Hennelly, D. P. Kelly, and T. J. Naughton, "Reconstruction algorithms applied to in-line Gabor digital holographic microscopy," Opt. Commun. **283**, 903-909 (2010).
56. C. Trujillo, P. Piedrahita-Quintero, and J. Garcia-Sucerquia, "Digital lensless holographic microscopy: numerical simulation and reconstruction with ImageJ," Appl. Opt. **59**, 5788-5795 (2020).
57. S. Shin, K. Kim, K. Lee, S. Lee, and Y. Park, "Effects of spatiotemporal coherence on interferometric microscopy," Opt. Express **25**, 8085-8097 (2017).
58. J. Dohet-Eraly, C. Yourassowsky, A. El Mallahi, and F. Dubois, "Quantitative assessment of noise reduction with partial spatial coherence illumination in digital holographic microscopy," Opt. Lett. **41**, 111-114 (2016).